\documentclass[12pt]{iopart}
\usepackage{graphicx}

\usepackage{iopams}  
\usepackage{hyperref}
\hypersetup{
    colorlinks=true,
    linkcolor=black,
    filecolor=black,      
    urlcolor=black,
    citecolor = black,
    pdftitle={A Toy Model for the Auditory System (arXiv)},
    pdfpagemode=FullScreen,
}

\usepackage[dvipsnames]{xcolor} % To highlight edits with the command \textcolor{}{}

\begin{document}

\title[A Toy Model for the Auditory System]{A Toy Model for the Auditory System that exploits Stochastic Resonance}

\author{Francesco Veronesi$^1$\footnote{Author to whom any correspondence should be addressed.} and Edoardo Milotti$^2$} 

\address{$^1$ School of Industrial and Information Engineering, Politecnico di Milano, Piazza Leonardo da Vinci 32, 20133 Milano, Italy}

\address{$^2$ Department of Physics, University of Trieste, Via Alfonso Valerio 2, 34127 Trieste, Italy}

\eads{\mailto{francesco2.veronesi@mail.polimi.it} and \mailto{milotti@units.it}}
\vspace{10pt}

%\begin{indented}
%\item[]December 2021
%\end{indented}

\begin{abstract}
The transduction process that occurs in the inner ear of the auditory system is a complex mechanism which requires a non-linear dynamical description. In addition to this, the stochastic phenomena that naturally arise in the inner ear during the transduction of an external sound into an electro-chemical signal must also be taken into account. The presence of noise is usually undesirable, but in non-linear systems a moderate amount of noise can improve the system's performance and increase the signal-to-noise ratio. The phenomenon of stochastic resonance combines randomness with non-linearity and is a natural candidate to explain at least part of the hearing process which is observed in the inner ear.
In this work, we present a toy model of the auditory system which shows how stochastic resonance can be instrumental to sound perception, and suggests an explanation of the frequency dependence of the hearing threshold.

\end{abstract}

\vspace{1pc}

\vspace{1pc}
\noindent{\it  Citation details}: Francesco Veronesi and Edoardo Milotti 2022 \textit{Eur. J. Phys.} \textbf{43} 025703. Version of Record DOI: \url{https://doi.org/10.1088/1361-6404/ac4431}

% This is the version of the article before peer review or editing, as submitted by an author to European Journal of Physics. IOP Publishing Ltd is not responsible for any errors or omissions in this version of the manuscript or any version derived from it. The Version of Record is available online at \url{https://doi.org/10.1088/1361-6404/ac4431}.

%\submitto{\EJP}

\section{Introduction}

Mathematical modeling in biophysics is notoriously difficult, because the majority of biological systems cannot be subdivided into hierarchically separated subsystems. The internal correlations and non-linear interactions are often so strong that the reductionist approach that is so successful in physics cannot be applied to biology \cite{novikoff1945concept,emmeche1997aspects,lobo2008biological}. Still, in some fortunate cases, simple physical models can account for the main observed features. For example, in 1977 Edward Purcell published a beautiful, seminal paper under the title ``Life at low Reynolds number'' that explained in simple terms the physical reasons underpinning the evolutionary development of some aquatic organisms \cite{Purcell1977}. 
This was followed a few years later by a similarly styled paper ``The efficiency of propulsion by a rotating flagellum'' \cite{purcell1997efficiency} which further extended the considerations of the 1977 paper, again with simple and deep physical arguments. Other notable contributions of physics to biophysics that stand out for their simplicity and depth can be found, e.g., in the fields of biomechanics \cite{lin1982fundamentals} and biophysical noise processes (see, e.g, \cite{adam1968reduction,berg1977physics,berg1985diffusion}, and for a modern perspective the beautiful book by William Bialek \cite{bialek2012biophysics}).

\medskip 

Here we try to follow these important leads while focusing on the complexity of the auditory system. The dynamical models used to describe the auditory systems are not analytically solvable, and the approximations used to predict the system's behaviour may compromise the overall reliability of the solutions. The intrinsic stochasticity of the underlying biological processes adds another layer of complexity \cite{3signore, Maoileidigh-IEHC}.

\medskip 

However, under appropriate conditions, the presence of noise in non-linear systems can improve their performance \cite{McDonnel}, in particular signal detection can benefit from noise and display an enhancement of the signal-to-noise ratio. This is the result of the phenomenon known as \textit{stochastic resonance}, first introduced by R. Benzi \etal \cite{Benzi} in 1981, and which was initially used to model the switching behaviour of the Earth climate that leads to the ice ages \cite{benzi1982stochastic}. Since its introduction stochastic resonance has been applied to a variety of fields, like, e.g.,  logic gates \cite{zhang2018realizing, zhang2017effect, zhang2018adaptive}, with extensions as far reaching as biophysics, see, e.g., \cite{wang2017array} which applies the concept to genetic networks.

\medskip 

In the context of the hearing system, stochastic resonance has been invoked as an explanation of tinnitus \cite{schilling2021stochastic} or to describe the sensation of pitch \cite{Martignoli-pitch}, thanks to the fact that it is compatible with neural models \cite{Moss+Ward+Sannita} and their threshold-like behaviour. 

\medskip 

In this paper we describe a simple model of the auditory system which is based on stochastic resonance, as defined in \cite{Gingl-Kiss-Moss, Gingl-Kiss-Moss-grafici}, that recreates to a good approximation the equal-loudness contours near the hearing threshold. The simplicity of the approach makes it well-suited as an introduction for BSc and MSc Physics students both to stochastic resonance and to the auditory system.

To make the paper self-contained, it starts with a brief introduction to the auditory system (section 2) and to stochastic resonance (section 3), followed by the description of the model. In section 4 we demonstrate that the model provides a good qualitative description of the equal-loudness curves. Finally we place the results in a wider context in the concluding section.

\section{Brief Overview of the Auditory System}\label{section:1}

\begin{figure}[ht]
	\includegraphics[width=0.4\textwidth]{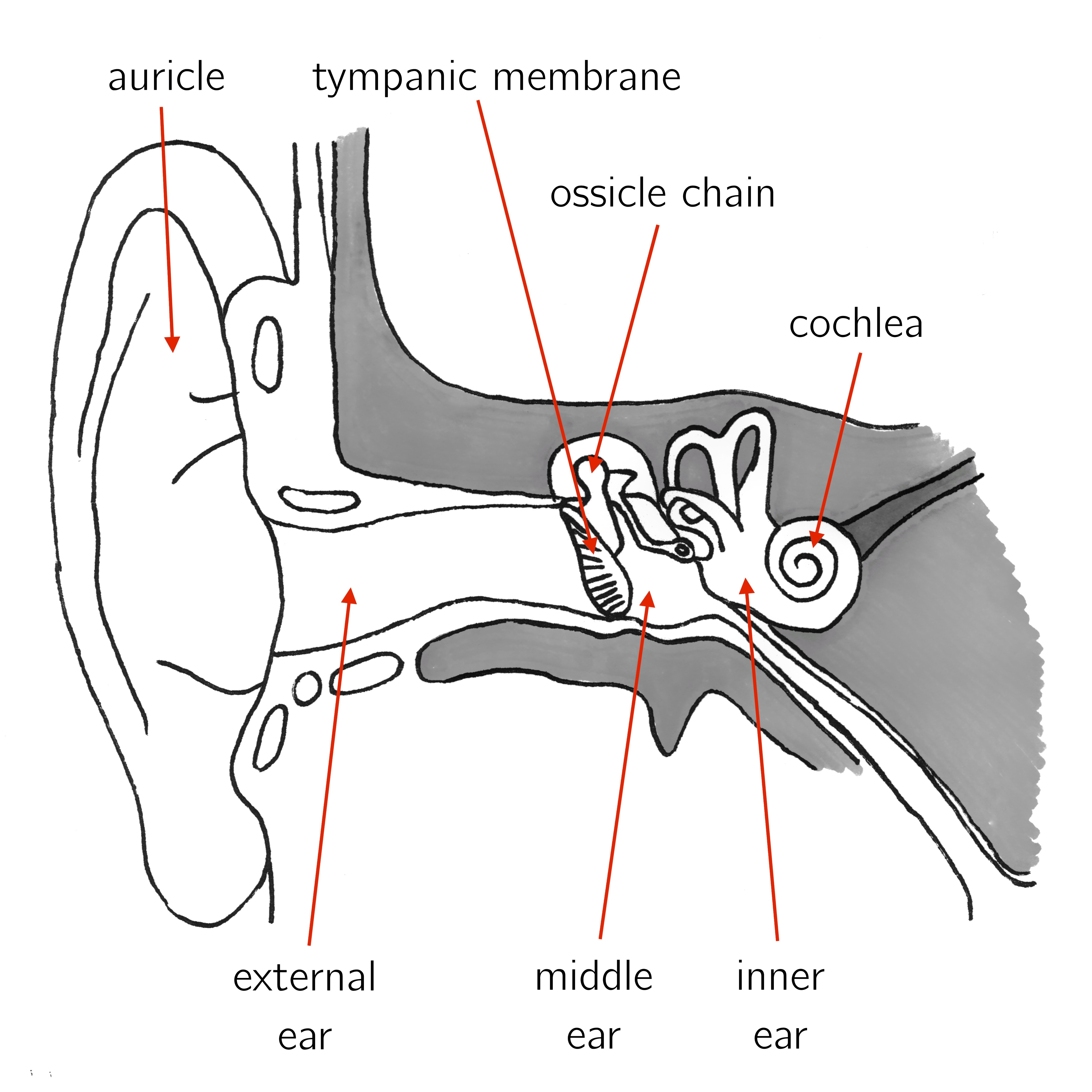}
	\centering
	\caption{Peripheral view of the auditory system and its major parts.}\label{fig:AuditorySystem}
\end{figure}

The human auditory system is a sensory organ composed of the outer (or external) ear, the middle ear, the inner ear, and the central auditory nervous system, whose overall function is to perceive and process sounds. The first two elements of the auditory system that are involved in the process, as shown in \Fref{fig:AuditorySystem}, are the outer and the middle ear. \medskip

%%%%%%%%%%%%%%%%%%%%%%%%%%%%%%%%%%%%%%
%%%%%%%%%%%%%%%%%%%%%%%%%%%%%%%%%%%%%%
%%%%%%%%%%%%%%%%%%%%%%%%%%%%%%%%%%%%%%
%%%%%%%%%%%%%%%%%%%%%%%%%%%%%%%%%%%%%%

\begin{figure}[ht]
  \centering
  \begin{minipage}{0.48\textwidth}
    \hspace*{1.8cm}\includegraphics[width=.9\textwidth]{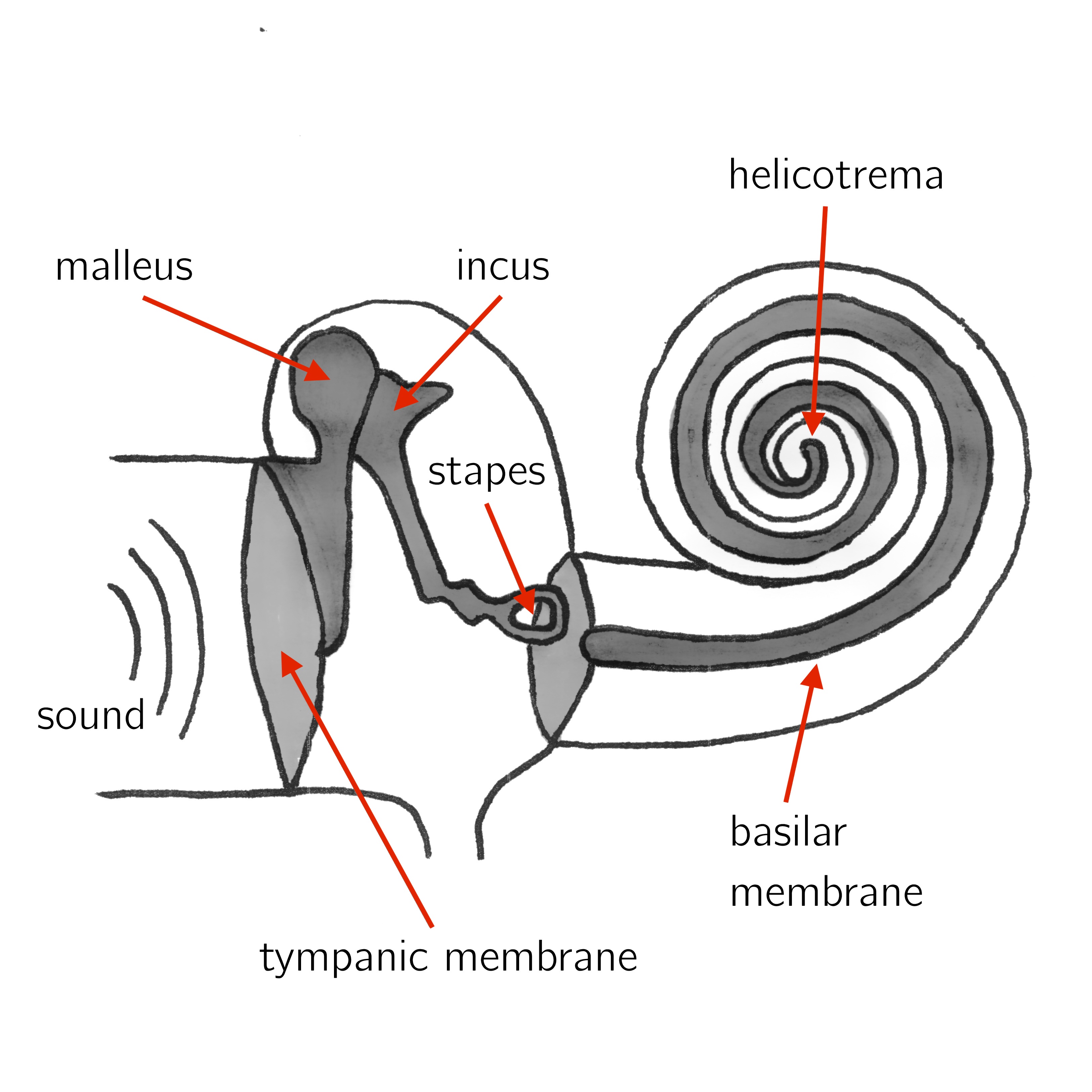}
    \caption{Schematic view of middle and inner ear. The basilar membrane divides the cochlea into two distinct compartments.}\label{fig:MidInnerEar}
  \end{minipage}
  \hfill
  \begin{minipage}{0.48\textwidth}
    \hspace*{1.8cm}\includegraphics[width=.9\textwidth]{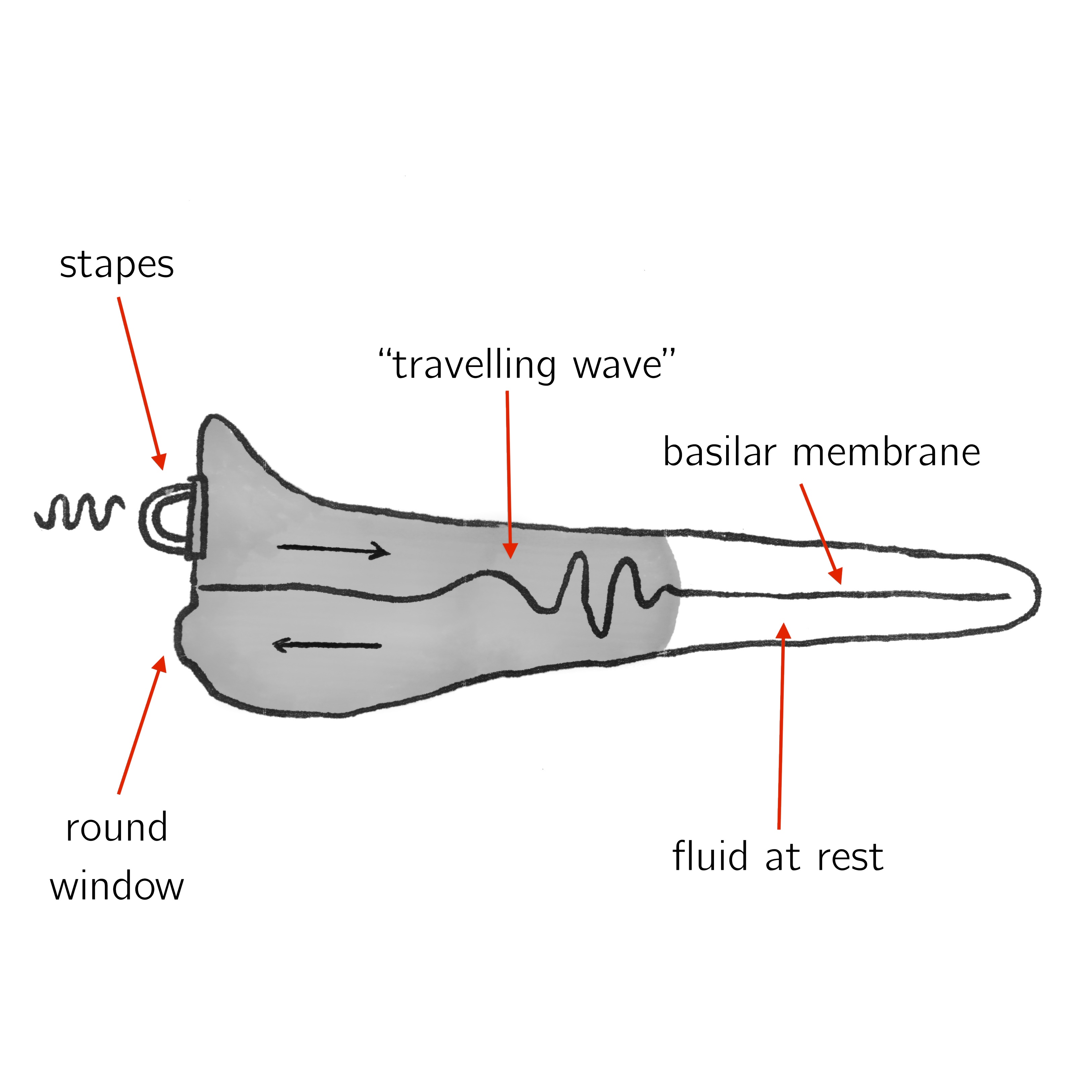}
    \caption{Representation of a wave propagating within the cochlea.}\label{fig:InnerEar}
  \end{minipage}
\end{figure}

The outer ear consists of the auricle and the ear canal. The former gathers and channels incident sound waves into the latter. Due to its relatively small size (small compared to the wavelengths of audible sounds), the auricle ensures its optimal operation point in the middle-high frequency region \cite{3signore}. At the center of the auricle we find the ear canal, a soft and rough body approximately cylindrical in shape. Its length varies according to age, gender and to genetic factors related with the subject. The ear canal acts like a resonant band-pass filter, with a resonant frequency in the range from 2~kHz to 3~kHz \cite{3signore}.\\
Once the sound wave, properly conveyed by the auricle and the ear canal, reaches the eardrum, its acoustic energy is converted into mechanical energy by the middle ear organs. The middle ear acts as an impedance between the outer ear (filled with air) and the inner ear (mainly filled with perilymph). The energy coming from the outer ear, as shown in \Fref{fig:MidInnerEar}, causes the vibration of the tympanic membrane which, in the middle ear, is transferred first to three small bones (malleus, incus and stapes) and then to the fluid that fills the anterior part of the cochlea, the main organ of the inner ear \cite{Mariola-Hearing}.

%%%%%%%%%%%%%%%%%%%%%%%%%%%%%%%%%%%%%%
%%%%%%%%%%%%%%%%%%%%%%%%%%%%%%%%%%%%%%
%%%%%%%%%%%%%%%%%%%%%%%%%%%%%%%%%%%%%%
%%%%%%%%%%%%%%%%%%%%%%%%%%%%%%%%%%%%%%

The cochlea is a spiral-shaped canal composed of several tunnels, each filled with a specific fluid (endolymph, perilymph). The motion of the stapes is transferred to those fluids through the oval window and then  absorbed by the basilar membrane, the prime structural element of the cochlea. The basilar membrane, depending on how it oscillates when excited by the pressure waves of the liquid (\Fref{fig:InnerEar}), is known to ensure sound-intensity and frequency encoding. In fact, the basilar membrane vibrations within the cochlea and the stimulation of its receptors, called hair cells, are converted into electro-chemical signals that reach the brain through the auditory nerve \cite{3signore}.  

The main theory that attempts to explain how the cochlea is able to encode the frequency of a signal is the \textit{place theory of hearing} \cite{3signore}. It assumes that a certain frequency is encoded by the position (place) along the basilar membrane where the amplitude of the vibration produced by the acoustic stimulus is at a maximum.\medskip

Moreover, according to the theory, each hair cell reacts to all frequencies stimuli but with distinct threshold values. 
Place theory does not explain how sound intensity is perceived and encoded. Current understanding suggests it is affected by \cite{3signore}
\begin{itemize}
    \item the number of hair cells that respond simultaneously to the same stimulus (since a high intensity sound stimulates a large number of hair cells);
    \item spontaneous activity of nerve fibers, which adds an additional degree of accuracy.
\end{itemize}

\begin{figure}[ht]
	\includegraphics[width=0.5\textwidth]{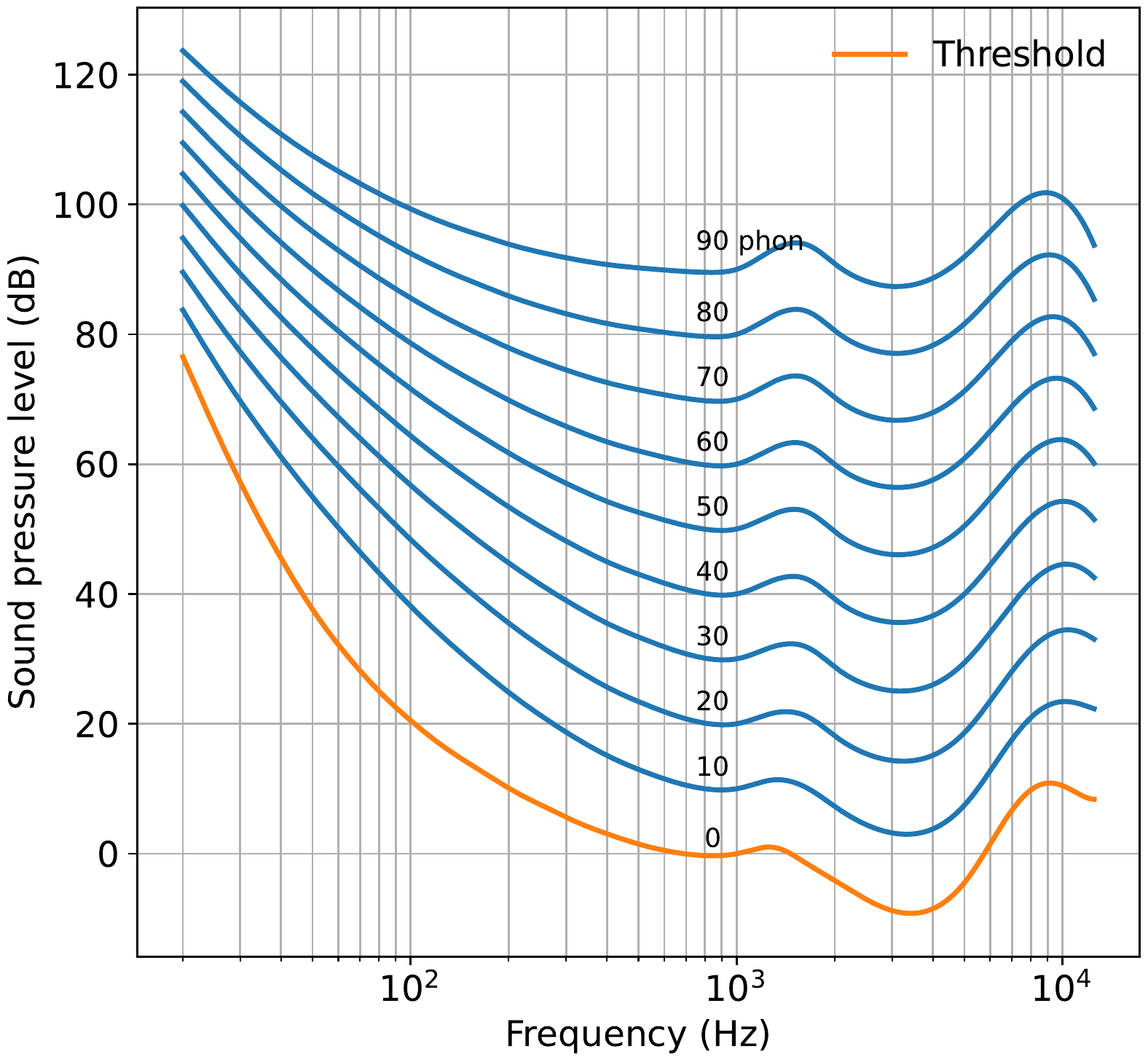}
	\centering
	\caption{Equal-loudness contours, as defined by standard 226 of the International Organization for Standardization (ISO) \cite{ISO-226:2003}.
	The basis of the equal-loudness contour is the phon, a unit of loudness that represents the dB Sound Pressure Level (SPL) necessary for a tone to elicit the same loudness as a 1000~Hz reference tone.}\label{fig:isofoniche}
\end{figure}

Graphically, sound-intensity perception is represented by equal-loudness contours, see \Fref{fig:isofoniche}. 
It is worth noting the presence of two minima (i.e., sensitivity maxima), one at a frequency just below 4~kHz and the other one at about 12~kHz (the values are similar to the resonance frequencies of the ear canal), and the behaviour at low frequencies ($<$ 500~Hz).\medskip

%%%%%%%%%%%%%%%%%%%%%%%%%%%%%%%%%%%%%%
%%%%%%%%%%%%%%%%%%%%%%%%%%%%%%%%%%%%%%
%%%%%%%%%%%%%%%%%%%%%%%%%%%%%%%%%%%%%%
%%%%%%%%%%%%%%%%%%%%%%%%%%%%%%%%%%%%%%

Among all the aspects that have emerged in this overview of the auditory system we emphasize the fact that the inner ear is a nonlinear dynamic system with threshold operation, whose transduction process (responsible for the transformation of an acoustic signal into an electro-chemical one) could be enhanced by the internal noise related to the spontaneous activity of hair-cell neurons.

\section{Introduction to Stochastic Resonance}

The term \emph{noise} describes random fluctuations or perturbations~\cite{McDonnel}, that introduce irregularities in physical signals~\cite{TWellens}. In systems with linear or weakly nonlinear dynamics an increase in noise intensity leads to a reduction of the signal-to-noise ratio (SNR), defined as the ratio between the mean signal power and the mean noise power, expressed in dB\footnote{Our choice to use the SNR as the proper figure of merit fits well with the convention used for equal loudness curves such as those shown in figure \ref{fig:isofoniche} and with the usual elementary treatments of stochastic resonance. For completeness, we remark that in recent years, when discussing the auditory system, an ever increasing emphasis is placed on its information theoretic properties \cite{hebert2002detection,overath2007information}.}. For a sinusoidal signal at a specific frequency $f_s$, the SNR can be evaluated from the respective power spectra densities $S$ and $S_N$ at the same frequency:
\begin{equation}\label{eq:SNR}
    SNR = \frac{S(f_s)}{S_N(f_s)}
\end{equation}

Surprisingly, in nonlinear systems there are circumstances where the presence of noise can lead to an increase of the SNR~\cite{Nacamara}: this is the phenomenon of \emph{stochastic resonance}. One can loosely interpret stochastic resonance as ``randomness that makes nonlinearity less detrimental to a signal'' \cite{McDonnel}.

Stochastic resonance was first introduced by R. Benzi~\cite{Benzi} at the NATO International School of Climatology~\cite{McDonnel}, where it was proposed as a possible explanation of some observed recurrences (approximately every 100 000 years) in the ice ages of the last 700 000 years~\cite{Nacamara,TWellens}. This phenomenon -- although not a real resonance -- was given the name of stochastic resonance because the signal-to-noise ratio (SNR) assumes its maximum value when the intensity of the input noise is ``tuned'' to a specific value~\cite{McDonnel, Nacamara}.

\begin{figure}[ht]
  \centering
  \begin{minipage}{0.48\textwidth}
    \hspace*{0.6cm}\includegraphics[width=1.1\textwidth]{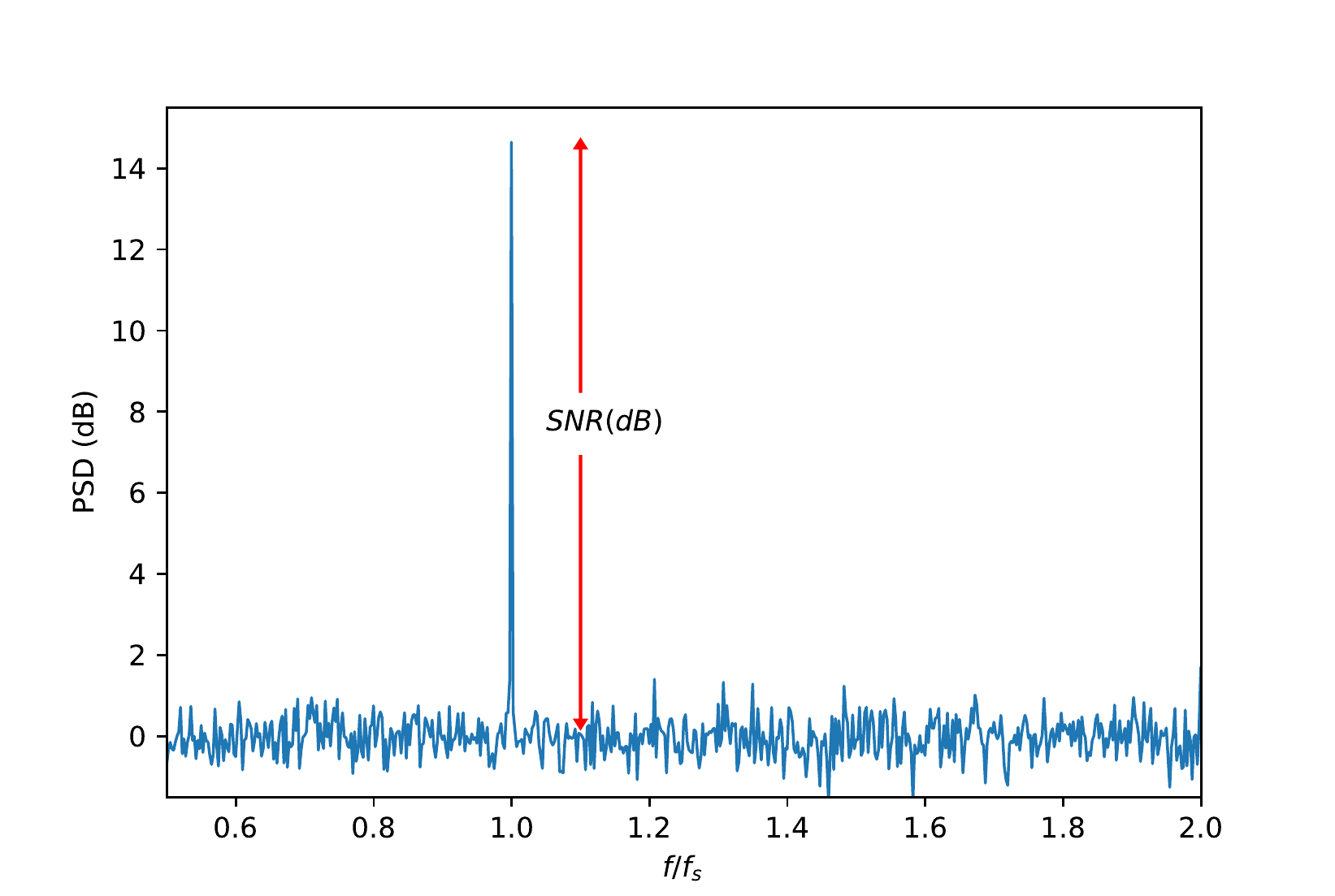}
    \caption{Power spectral density (PSD) of a generic system which exhibits stochastic resonance. The peak height of the signal is used to compute the signal-to-noise ratio (SNR) at the frequency $f_s$ of the signal. In this simulation a zero-mean white noise has been used. Note the representation in dB. }\label{fig:PSDtheor}
  \end{minipage}
  \hfill
  \begin{minipage}{0.48\textwidth}
    \hspace*{1.5cm}\includegraphics[width=.9\textwidth]{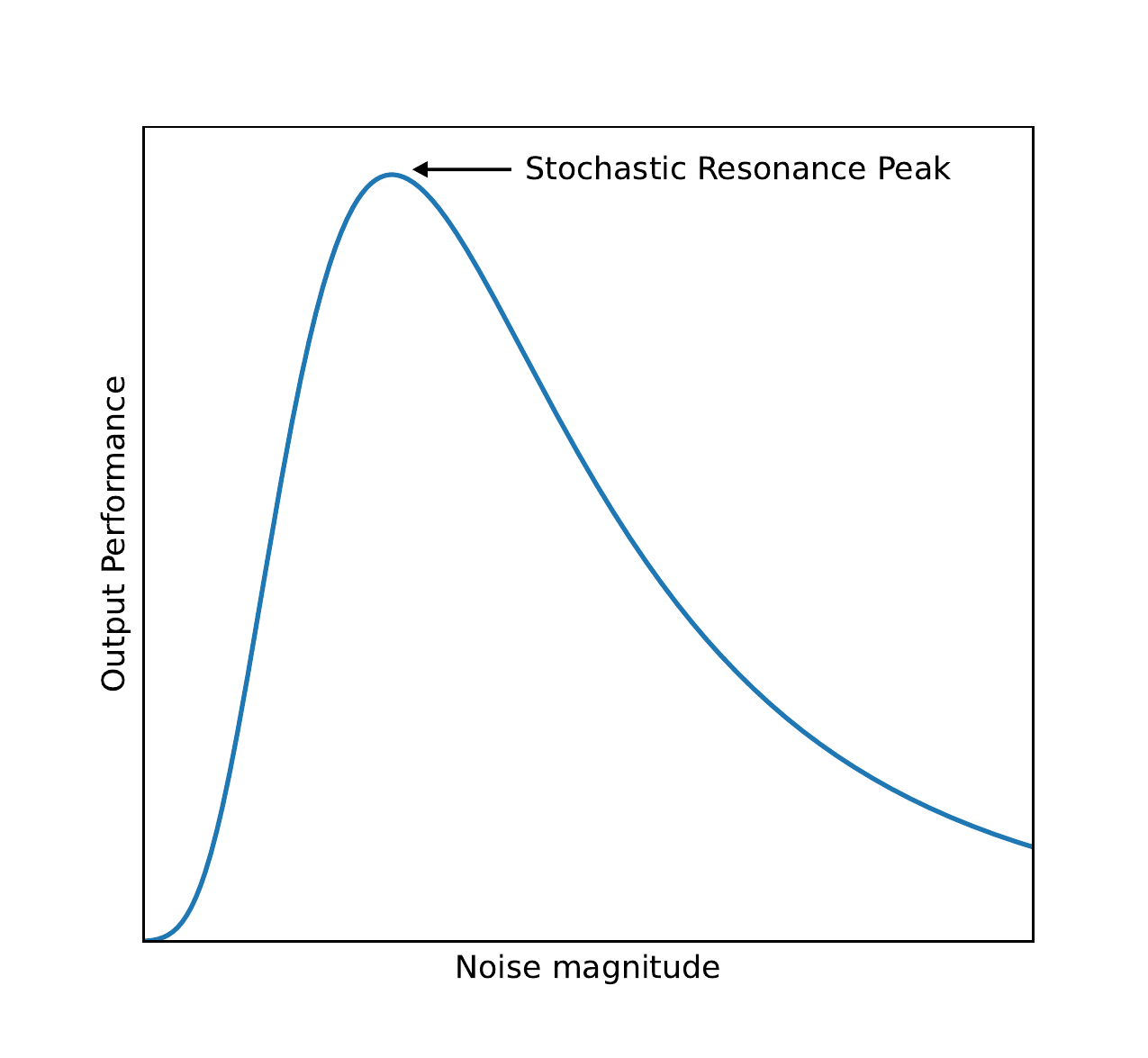}
    \caption{The plot shows the typical SNR -- denoted here as Output Performance -- vs. noise magnitude for a system that exhibits stochastic resonance.}\label{fig:SNRtheor}
  \end{minipage}
\end{figure}

Given the ubiquity of noise in nature -- and more specifically in biophysical and physiological contexts, where nonlinearity is widespread -- this property has prompted searches for the existence and the manifestation of stochastic resonance in neural and sensory models. 
The ``cooperation'' that arises between  signal and  noise introduces a coherence in the system that is quantified very conveniently by means of the power spectral density (PSD) associated to the system~\cite{TWellens, Tuning-Into-Noise}. In fact, if stochastic resonance is realized between noise and a pure sinusoidal signal of frequency $f_s$, then the power spectrum displays a peak at frequency $f_s$ (see \Fref{fig:PSDtheor}). The height of this peak is both frequency- and noise-intensity-dependent~\cite{TWellens, Tuning-Into-Noise}.
The dependence of SNR on noise amplitude also exhibits a similar behavior (see \Fref{fig:SNRtheor}). Therefore, stochastic resonance is said to occur if this plot displays a maximum \cite{TWellens} or, equivalently, is characterized by an inverted U-shape \cite{Eduardo-Lugo}.  This type of trend -- typical of all types of stochastic resonance \cite{McDonnel} -- is considered the hallmark of the effect \cite{Tuning-Into-Noise}.

%%%%%%%%%%%%%%%%%%%%%%%%%%%%%%%%%%%%%%
%%%%%%%%%%%%%%%%%%%%%%%%%%%%%%%%%%%%%%
%%%%%%%%%%%%%%%%%%%%%%%%%%%%%%%%%%%%%%
%%%%%%%%%%%%%%%%%%%%%%%%%%%%%%%%%%%%%%

For historical reasons, it is customary to distinguish between dynamical stochastic resonance and non-dynamical stochastic resonance. In fact, the original stochastic resonance presented by R. Benzi \etal \cite{Benzi} was a phenomenon that occurred only in bistable or multistable dynamical systems \cite{Gingl-Kiss-Moss, Gingl-Kiss-Moss-grafici}, whose definition required the verification of precise conditions. Moreover, this definition made the word stochastic resonance inappropriate for nonlinear systems where the nonlinearity was due solely to a ``static threshold'' \cite{McDonnel}. For these reasons, nowadays, it is usual to differentiate between the original dynamical stochastic resonance and ``static'', or non-dynamical, stochastic resonance, despite the fact that both types exhibit the same properties presented so far.
%For these reasons, nowadays, it is usual to differentiate between dynamical stochastic resonance and the original ``static'', or non-dynamical, stochastic resonance, despite the fact that both types exhibit the same properties presented so far.

In this paper we deal only with non-dynamical stochastic resonance, giving the opportunity, to those interested, to delve into the dynamical one by consulting, for example, the articles by T. Wellens et al. (2004) \cite{TWellens}, by B. McNamara and K. Wiesenfeld (1989) \cite{Nacamara} and by A. B. Bulsara and L. Gammaitoni (1996) \cite{Tuning-Into-Noise}.%\medskip

\begin{figure}[ht]
	\hspace*{1.5cm}\includegraphics[width=0.95\textwidth]{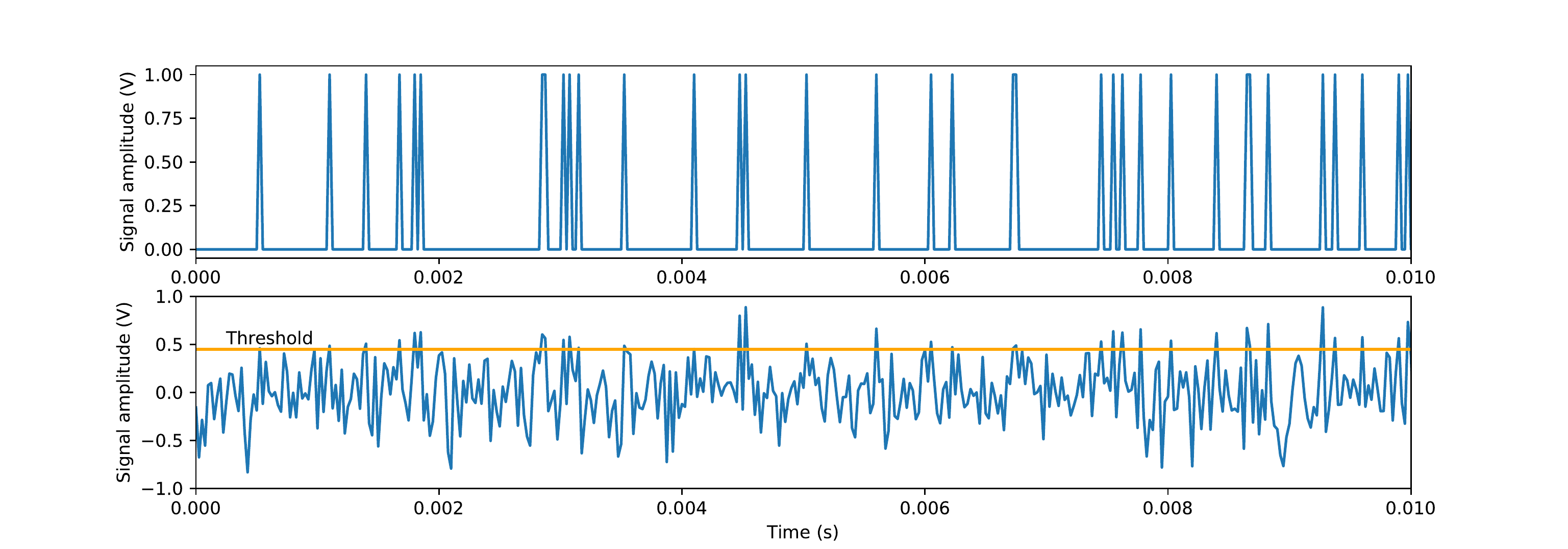}
	\centering
	\caption{Input (lower panel) and output (upper panel) representation of an asymmetric Level Crossing Detector system. Here the threshold is represented by the orange line. Whenever the input -- given by the signal plus noise -- exceeds the threshold, a short pulse of arbitrary amplitude is added to the output time series.}\label{fig:SpikeTrain}
\end{figure}

A system that exhibits stochastic resonance is said to be ``static'' when the nonlinear perturbations to which it is subjected, and which alter the nature of the input signal, are not governed by temporal differential equations, but by simple dynamical rules that produce an output signal related to an instantaneous value assumed by the input signal \cite{McDonnel}. The simplest static system in which non-dynamical stochastic resonance occurs, as shown in the lower plot of \Fref{fig:SpikeTrain}, consists solely of a threshold detector and is called \textit{Level Crossing Detector} (LCD) \cite{Gingl-Kiss-Moss, Gingl-Kiss-Moss-grafici}.

LCDs base their operation on the following rule: whenever the input given by the sum of signal plus noise crosses the threshold, a narrow pulse of arbitrary amplitude is reported in the time series, as shown in \Fref{fig:SpikeTrain} for a pure sinusoidal signal. Depending on whether one chooses to subject the system to a single threshold (usually positive) or two (one positive and one negative), the LCD system is called asymmetric or symmetric, respectively. 
%In cases where the amplitude of the sinusoidal signal is comparable with the threshold value, the difference between the two systems is in the resonance of the harmonics: in symmetrical systems, in fact, increasing the input signal strength is not sufficient to bring out the signal harmonics in the PSD graph, an aspect that occurs, instead, in asymmetrical systems \cite{Gingl-Kiss-Moss, Gingl-Kiss-Moss-grafici}.

%%%%%%%%%%%%%%%%%%%%%%%%%%%%%%%%%%%%%%
%%%%%%%%%%%%%%%%%%%%%%%%%%%%%%%%%%%%%%
%%%%%%%%%%%%%%%%%%%%%%%%%%%%%%%%%%%%%%
%%%%%%%%%%%%%%%%%%%%%%%%%%%%%%%%%%%%%%

After the publication of first work on stochastic resonance in neuronal models in 1991, which was soon followed by experimental observations in 1993 by studying the functioning of crayfish's mechanoreceptors \cite{McDonnel, Moss+Ward+Sannita}, the presence of stochastic resonance has been theorized in various biological contexts \cite{schilling2021stochastic, wiesenfeld1995stochastic}. To this day, it is still debated whether it can play a role in neuroscience, and in particular in the sensory functions of touch, hearing and vision \cite{Moss+Ward+Sannita}.\medskip

Up until now, it is not yet clear whether neurons do make use of stochastic resonance \cite{wiesenfeld1995stochastic}, and the evidence that they actually exploit it is only indirect \cite{McDonnel}. In fact, in most experimental settings the noise input to the sensory receptors or neurons comes from external sources. For this reason, any manifestation of stochastic resonance only allows to deduce that sensory cells are nonlinear dynamical systems that could benefit from the presence of intrinsic noise in neural processing \cite{McDonnel}. Despite this, stochastic resonance remains a phenomenon that is compatible with several neural models and some theories of neural processing \cite{Moss+Ward+Sannita}. Indeed,  neurons are known to be intrinsically noisy, with a behavior that is similar to threshold systems \cite{wiesenfeld1995stochastic}: whenever a certain internal threshold is exceeded a neuron ``fires'', generates a ``nerve impulse'' (action potential) and returns to the resting state waiting for a new supra-threshold event \cite{Eduardo-Lugo}. 
It is clear that dynamical stochastic resonance could play an important role in the functioning of neurons or sensory cells \cite{Martignoli-pitch, Eduardo-Lugo, Krauss-upregulation, FitzHug-neuron-model}. However, studying such systems in the non-dynamical approximation makes the discussion simpler and equally valid \cite{schilling2021stochastic, Moss+Ward+Sannita, Tuning-Into-Noise}. %, if we take into account the limited degree of knowledge that we still have of systems such as the human brain, the visual system and the auditory system \cite{schilling2021stochastic, Moss+Ward+Sannita, Tuning-Into-Noise}.

It is interesting to note that a LCD produces both detection and a kind of pulse-train encoding similar to that found in dedicated electronic circuits \cite{yen2009integrated}: the amplitude of sub-threshold stimuli is encoded into the frequency of threshold crossings. The incoming stimuli can be sub-threshold and therefore undetectable. If noise is added to the stimulus, threshold crossing occurs with higher probability when the stimulus is close to the threshold. The resulting spike train, despite being ``noisy'', contains a large part of the information carried by the sub-threshold signal. If one compares this situation with that in which noise is the only signal present, whereby the threshold crossing occurs randomly, one deduces that the extra information that is found in the spike train generated by a non-stochastic signal ensures that the sub-threshold stimulus is well-characterized.

We can therefore say that noise activates a random sampling of the stimulus. %This means that the largest values assumed by the noise are favoured to exceed the threshold and thus to provide a ``sample'' of the sub-threshold signal at a certain instant of time. 
Therefore, for good information transmission, the ``sampling rate'' should be greater than the frequency of the sub-threshold signal. A convenient measure of the quality of the output signal (pulse train) from the threshold system, and thus of how well it is able to represent the sub-threshold signal, is precisely the SNR, which can be used to find the optimal threshold level for a given noise intensity \cite{Jung-spike-train}.%\medskip

The auditory system, as seen in the previous section, is very complex and is composed of several nonlinear sub-structures. Since noise is ubiquitous in the sensory systems \cite{Martignoli-pitch, Moss+Ward+Sannita}, it is clear that the auditory system could exploit, for its operation or in some of its parts, stochastic resonance \cite{Moss+Ward+Sannita}. Considerations of this kind have been studied and debated in several contexts \cite{schilling2021stochastic, Moss+Ward+Sannita, JHuang-speech, Huang-melody}. There is not a real consensus, e.g., Rufener \etal \cite{Rufener} carried out experiments by applying external noise and do not find an enhanced sound perception, however the application of external noise reduces the SNR in a well-tuned stochastic resonance system, and their results do not disprove the importance of stochastic resonance. 

\medskip

Here we focus only on the role that stochastic resonance can play in the transduction process that takes place in the inner ear, which involves the cochlea, the inner hair cells and the neurons of the auditory nerve. The signal detected by the cochlea and processed by the hair cells activates the neurons of the auditory nerve. At first glance, their extreme noisiness seems to hinder their ability to transmit precise sounds and acoustic signals (an ability that depends, in a decisive way, on exact timing and frequencies). However, stochastic resonance does help, and the presence of noise has beneficial effects \cite{schilling2021stochastic, Martignoli-pitch}. 

\section{Toy Model of the Auditory System}

In this section we present the basic features of a simple model of human hearing based on stochastic resonance. 
The main hypothesis behind the model is that stochastic resonance is a phenomenon continuously occurring in the human auditory system,  which provides a simple transduction mechanism. The main element of the model is a symmetrical LCD system, which reproduces the behavior of human hearing in the context of loudness perception for sounds close to the hearing threshold. \medskip

The input signal is the sum of a sinusoidal waveform and noise~\cite{Gingl-Kiss-Moss, Gingl-Kiss-Moss-grafici}:
\begin{equation}\label{eq:V-in}
   V_{in}(t) = \varepsilon\sin{(2\pi f_s t)} + G(t) 
\end{equation}
where $\varepsilon$ and $f_s$ are respectively the amplitude and frequency of the sub-threshold sinusoidal signal, while $G(t)$ is the noise that is added to the process. 

Unlike the LCD system presented in \cite{Gingl-Kiss-Moss, Gingl-Kiss-Moss-grafici}, the output signal $V_{out}(t)$ is equal to 0 when $V_{in}$ does not exceed the threshold and it is equal to the deviation between the signal and the threshold in the other cases:
\begin{equation}
\label{eq:THR-condition}
V_{out}(t) = \left\{ \begin{array}{rcl} 
     V_{in}(t) \mp \Delta U & \mbox{if} & \pm V_{in}(t) > \Delta U \\
     0 & \mbox{if} & \ |V_{in}(t)|\leq \Delta U
\end{array}\right.
\end{equation}

The choice of a symmetric LCD system such as the one defined by \Eref{eq:THR-condition} is, in our opinion, the most appropriate for a model that aims to simulate the threshold behavior of one or more neurons.
The resulting LCD is simulated by generating evenly spaced $V_{in}$ samples. We take the sampling rate, $f_c = 40$~kHz to cover the audible frequency band up to the 20~kHz Nyquist frequency. The total sampling time is $T=0.1$~s, so that signals with frequency $f_s<10$~Hz fail to successfully complete a cycle and must be rejected. Again, this choice is justified by the lower frequency threshold of human hearing at about 12~Hz \cite{12Hz}.

The signal of frequency $f_s$ combines, on its way to the auditory nerve, with various noise sources, some external, others internal, which together concur to produce stochastic resonance. In this toy model we choose white, Gaussian noise (zero mean and variance $D^2$). This choice, as discussed in \cite{Martignoli-pitch}, can be considered acceptable although it is not always plausible. 

\begin{figure}[ht]
  \centering
  \begin{minipage}{0.49\textwidth}
    \hspace*{0.7cm}\includegraphics[width=1.1\textwidth]{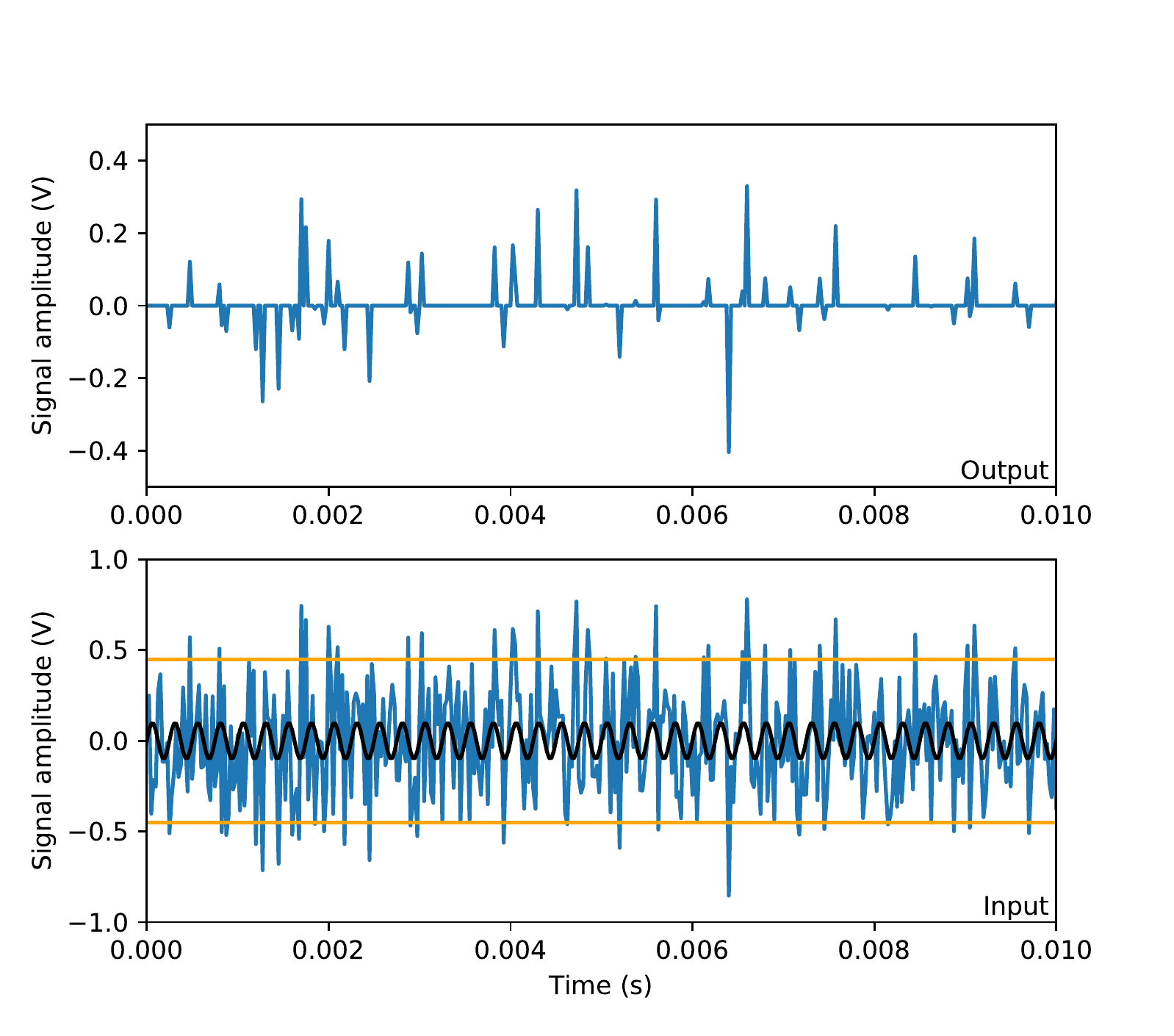}
    \caption{Graphical representation of the LCD system for $f_s=4$~kHz, $\Delta U = 0.45$~V, $D=0.3$~V and $\varepsilon=0.1$~V. \\Top: signal $V_{out}(t)$ as defined in \Eref{eq:THR-condition}; bottom: signal $V_{in}(t)$ (blue) as defined in \Eref{eq:V-in}, the thresholds $\pm \Delta U$ (orange) and the signal $\varepsilon \sin{(2\pi f_s t)}$ (black).}\label{fig:LCD-4000Hz}

  \end{minipage}
  \hfill
  \begin{minipage}{0.49\textwidth}
    \hspace*{.8cm}\includegraphics[width=1.1\textwidth]{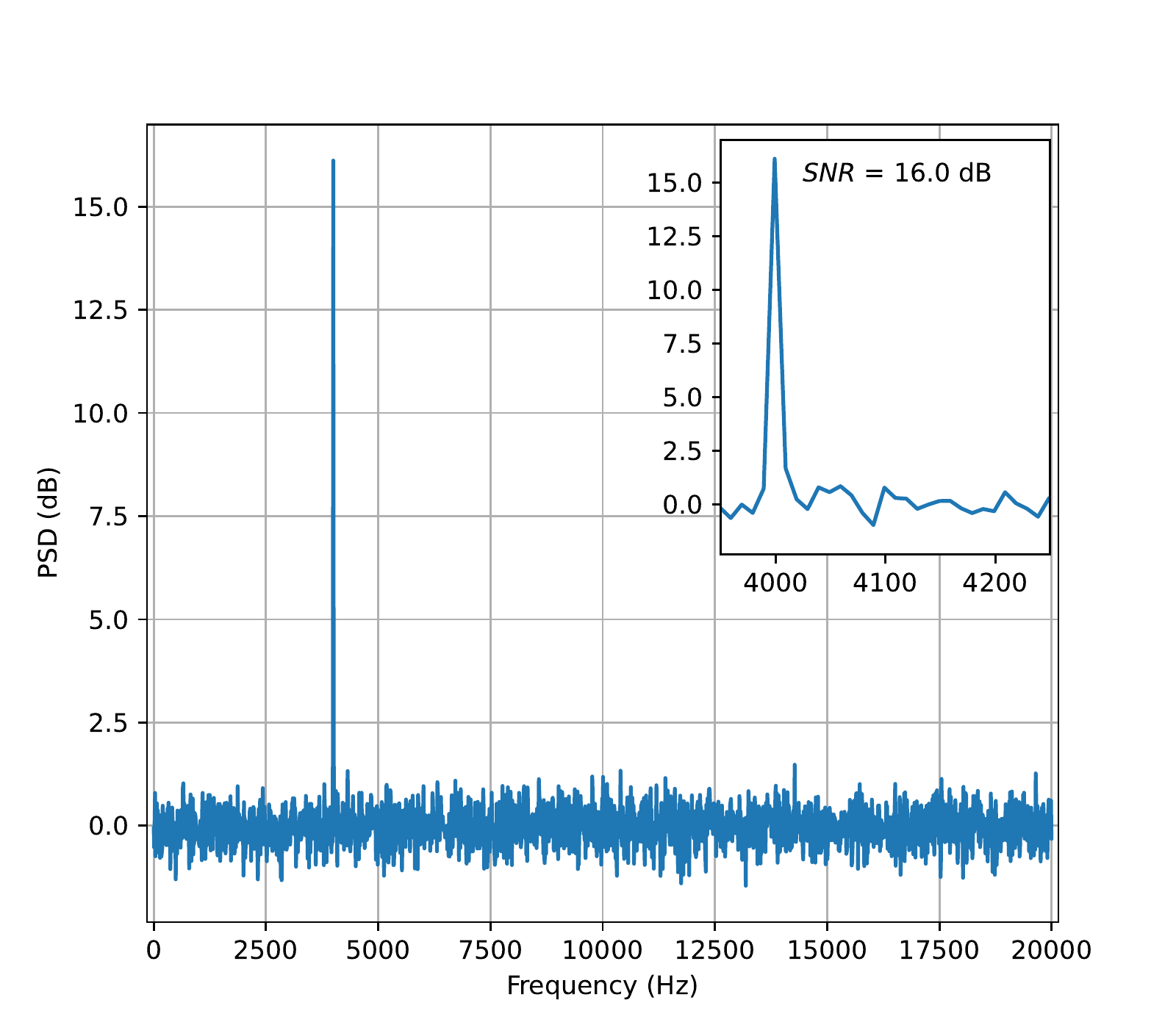}
    \caption{Power spectral density (PSD) of $V_{out}(t)$ as defined in \Eref{eq:THR-condition} for $f_s=4$~kHz, $\Delta U = 0.45$~V, $D=0.3$~V and $\varepsilon=0.1$~V. \\
    The dB representation of the PSD has been realised by choosing the average noise power as the reference parameter.\\
    %Signal-to-noise ratio (SNR) was computed using \Eref{eq:SNR}.
    }\label{fig:PSD-4kHz}
  \end{minipage}
\end{figure}

For the model we do not use physiological values. The values of the threshold $\Delta U$, the standard deviation $D$ of the white Gaussian noise and the amplitude of the sinusoidal signal $\varepsilon$, with which the simulations are performed (see \Fref{fig:LCD-4000Hz}), are chosen, for convenience, to be of the order of 100~mV\footnote{The resulting values are easier to read and the toy model -- by its nature -- is scale independent, so that the actual values do not matter here.}, and therefore two orders of magnitude larger than the typical values of the auditory system (mV) \cite{Eduardo-Lugo}. Accordingly, the amplitude $\varepsilon$ is chosen in such a way that the sinusoidal signal is always sub-threshold ($\varepsilon<\Delta U$).\\

The PSD of the output signal is estimated by taking the scalar average of the PSD computed with the FFT algorithm for a number of simulations (preferably $\gg$ 10). This approach reduces the dispersion in each frequency bin (see \Fref{fig:PSD-4kHz}) and provides a more accurate evaluation of the SNR. 

%%%%%%%%%%%%%%%%%%%%%%%%%%%%%%%%%%%%%%
%%%%%%%%%%%%%%%%%%%%%%%%%%%%%%%%%%%%%%
%%%%%%%%%%%%%%%%%%%%%%%%%%%%%%%%%%%%%%
%%%%%%%%%%%%%%%%%%%%%%%%%%%%%%%%%%%%%%

\begin{figure}%[th!]
  \centering
  \begin{minipage}{0.49\textwidth}
    \hspace*{0.6cm}\includegraphics[width=1.1\textwidth]{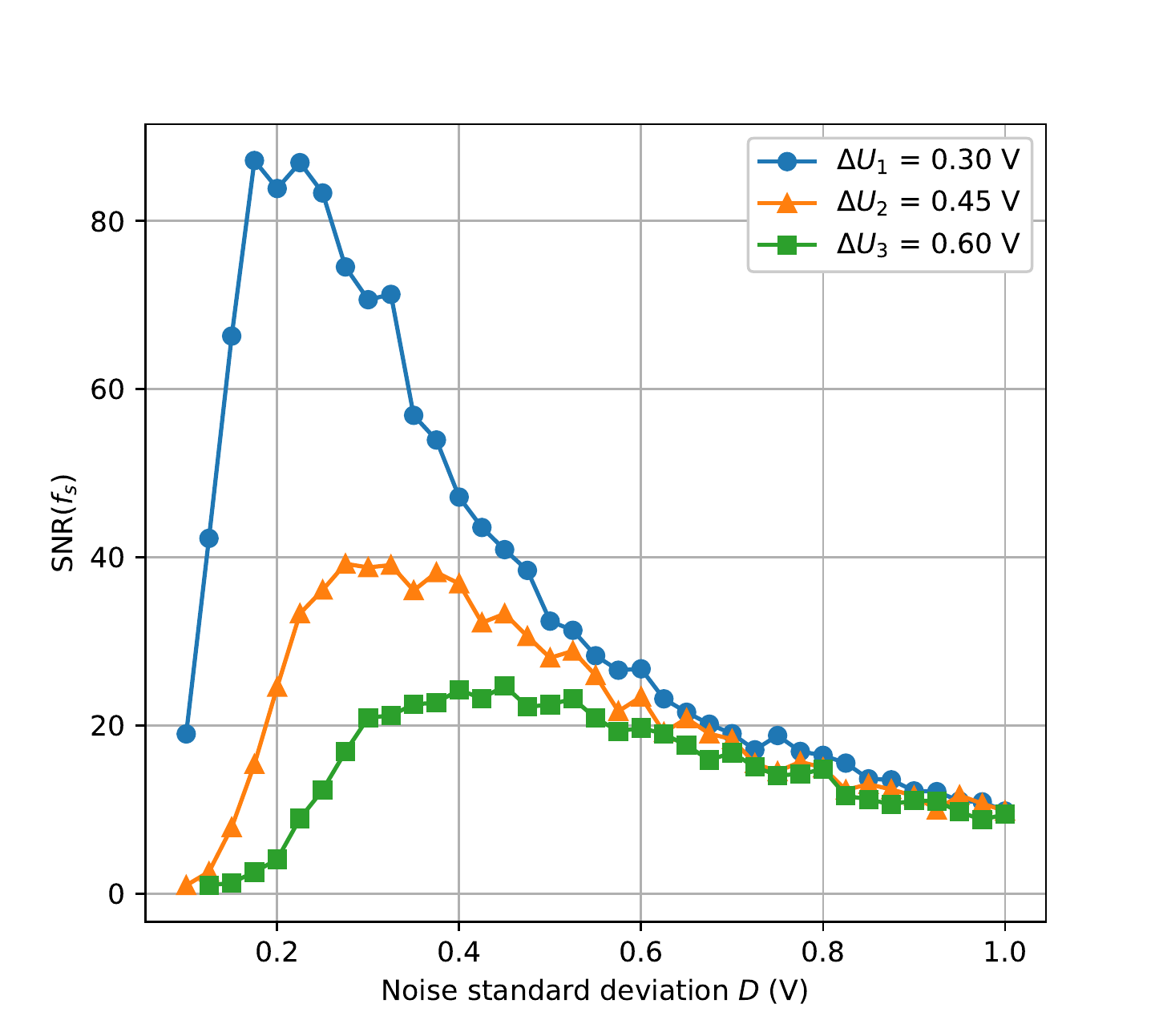}
    \caption{Signal-to-noise ratio (SNR) at the frequency $f_s=4$~kHz as a function of noise standard deviation for three different threshold values ($\Delta U_i$). We choose $\varepsilon=0.1$~V for the amplitude of the input sinusoidal signal.}\label{fig:SNR-4kHz}
  \end{minipage}
  \hfill
  \begin{minipage}{0.49\textwidth}
    \hspace*{0.6cm}\includegraphics[width=1.1\textwidth]{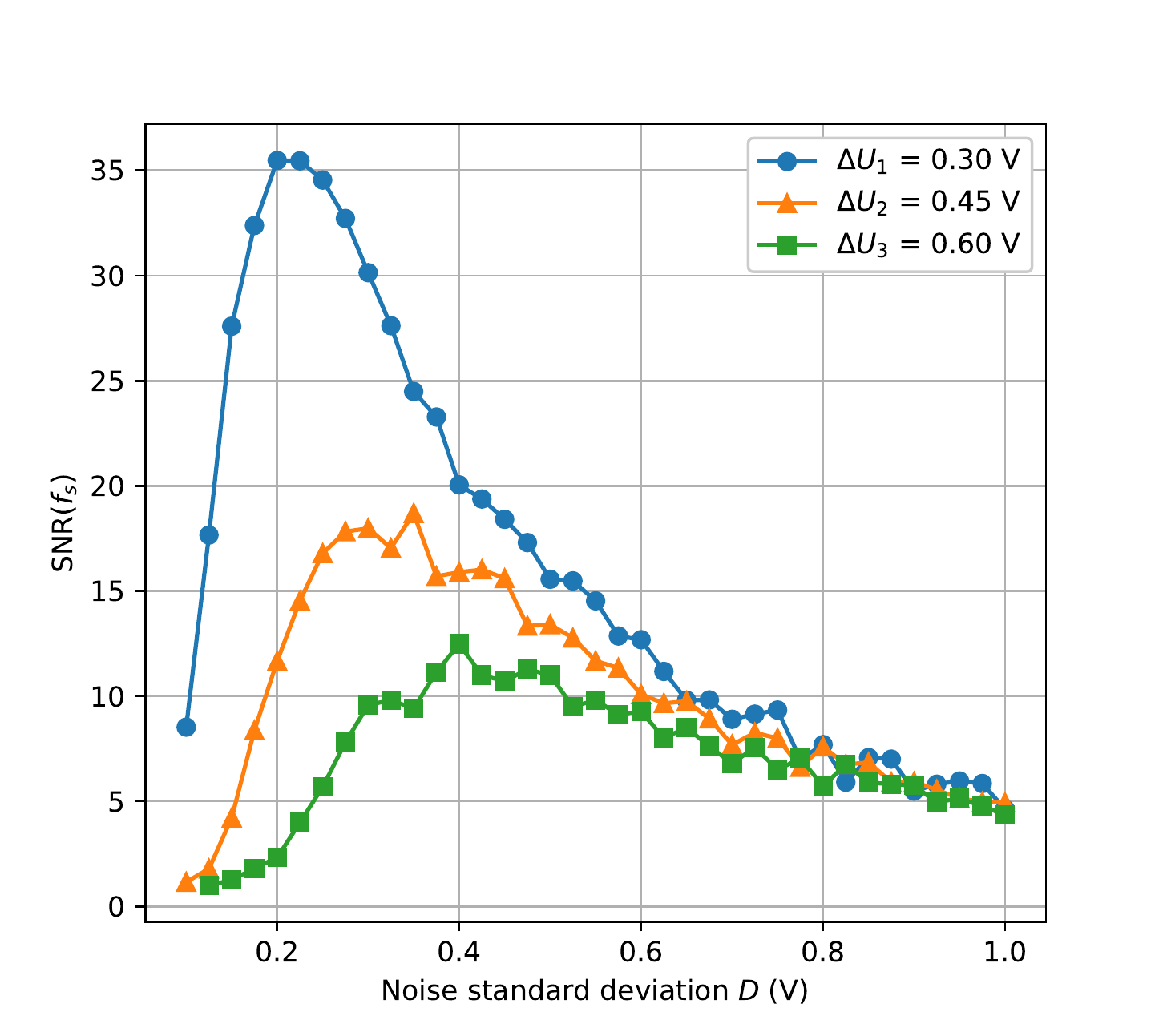}
    \caption{Signal-to-noise ratio (SNR) at the frequency $f_s=19$~kHz as a function of noise standard deviation for three different threshold values ($\Delta U_i$). We choose $\varepsilon=0.1$~V for the amplitude of the input sinusoidal signal.}\label{fig:SNR-19kHz}
  \end{minipage}
\end{figure}

Plotting the dependence of the SNR vs. the standard deviation of noise allows to find that stochastic resonance does occur in the system. This has been verified (see \Fref{fig:SNR-4kHz}, \Fref{fig:SNR-19kHz}) choosing values of $D$ between $D_\mathrm{min}= 0.1$~V and $D_\mathrm{max}= 1.0$~V and for three different threshold values: $\Delta U_0=0.30$~V, $\Delta U_1=0.45$~V, $\Delta U_2=0.60$~V. 

Observing the graphs of \Fref{fig:SNR-4kHz} and \Fref{fig:SNR-19kHz}, it can be noted that at high frequencies the signal-to-noise ratio is about half that at low frequencies. In LCD systems, optimal information transmission depends on the sampling frequency chosen to simulate the system. When the sinusoidal signal takes a time similar to $2\delta t = (f_c/2)^{-1}$ to complete a full cycle, the noise amplitude varies with a frequency similar to that of the signal. This condition increases the frequency at which the threshold is exceeded. This results in a signal that is noisier than desired (but still exhibits stochastic resonance) and thus in a lower signal-to-noise ratio than the one observable for $f_s\ll f_c/2$.

%%%%%%%%%%%%%%%%%%%%%%%%%%%%%%%%%%%%%
%%%%%%%%%%%%%%%%%%%%%%%%%%%%%%%%%%%%%%
%%%%%%%%%%%%%%%%%%%%%%%%%%%%%%%%%%%%%%
%%%%%%%%%%%%%%%%%%%%%%%%%%%%%%%%%%%%%%

The sensitivity of the human ear to loudness, as seen previously in \sref{section:1}, reaches a maximum in the medium-high frequency range (1~kHz -- 4~kHz), while it is lower at low frequencies ($<$ 0.2~kHz). This characteristic does not depend on the hair cells or on the physiology of the inner ear, but on the shape of the auditory canal, inside which the pure signal, mixed with external noise, propagates.
For this reason we assume that sound is filtered at low frequencies before reaching the cochlear membrane. Provided that the external noise is absent or negligible compared to the internal noise, in this toy model we add a high-pass filter to the LCD system that acts only on the sinusoidal signal, before combining with noise.
The first-order IIR high-pass filter we use has cut-off frequency at 30~Hz.

In order to produce equal-loudness contours with the available model and thus describe how the perception of sounds close to the threshold of hearing can occur, the following considerations were used:
\begin{itemize}
    \item assuming that the sound field consists of free progressive plane waves, the sound intensity $I_{out}$ depends on the sound amplitude $\varepsilon$:  $I_{out}\propto \varepsilon^2$. We take the intensity $I_0$ at 1~kHz as the reference intensity.
    
    \item we take the signal-to-noise ratio associated with the output signal of the LCD system as a measure of the  perceived sound intensity $I_{in}$. Thus, if we keep in mind that $SNR\propto \varepsilon^2$ \cite{Gingl-Kiss-Moss,wiesenfeld1995stochastic,RS-tutorialandupdate} for fixed values of noise intensity and threshold and we assume, by appealing the plasticity of the auditory system \cite{Irvine-Plasticity}, that the internal noise intensity is constantly optimal and therefore guarantees $I_{in}\propto \varepsilon^2$ even beyond the LCD \cite{schilling2021stochastic}, then, at the output of the LCD system, loudness of the sub-threshold signal is such that

\begin{equation}\label{eq:matta}
    I_{in}\propto SNR
\end{equation}

    Thanks to this, it is reasonable to conclude that the set of  $\varepsilon$ values that correspond to a constant signal-to-noise ratio defines a candidate equal-loudness contour.
\end{itemize}

%%%%%%%%%%%%%%%%%%%%%%%%%%%%%%%%%%%%%%
%%%%%%%%%%%%%%%%%%%%%%%%%%%%%%%%%%%%%%
\begin{figure}[ht]
	\hspace*{1.1cm}\includegraphics[width=0.8\textwidth]{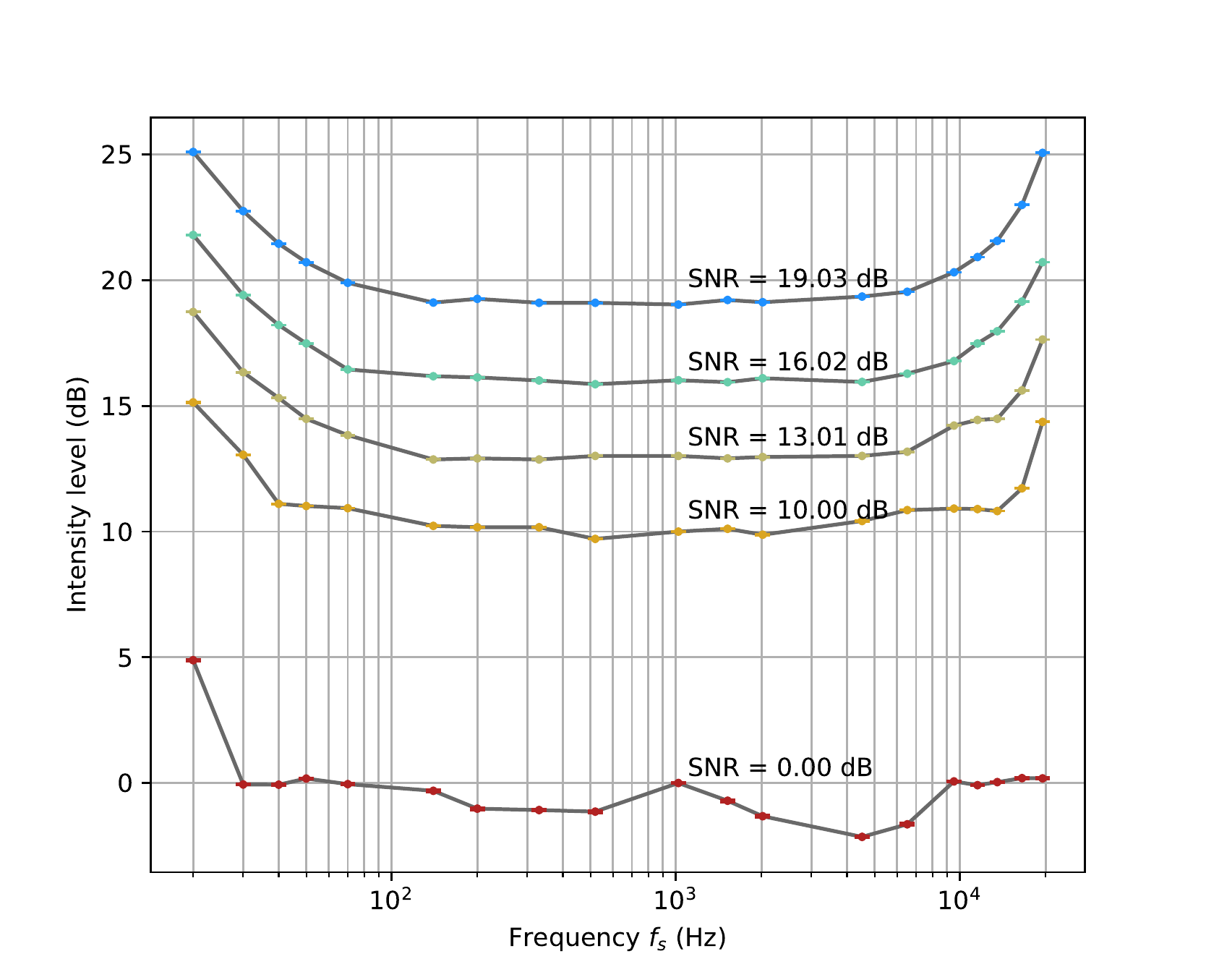}
	\centering
	\caption{Equal-SNR contours as a function of the external sound intensity level and the $f_s$ frequency of the sinusoidal signal.  For the simulations we have selected $\Delta U = 0.45$~V (threshold), $D=0.3$~V (noise standard deviation) and $f_{3dB}=30$~Hz (filter cut-off frequency).}\label{fig:EqualSNR}
\end{figure}
%%%%%%%%%%%%%%%%%%%%%%%%%%%%%%%%%%%%%%
%%%%%%%%%%%%%%%%%%%%%%%%%%%%%%%%%%%%%%

Using these considerations we find the equal-SNR contours shown in \Fref{fig:EqualSNR}.  
The behaviour at low frequencies is determined by the high-pass IIR filter. The frequencies $f_s$ of the sinusoidal signal are chosen in such a way that they are equally spaced in the logarithmic scale graph and do not produce \textit{scalloping loss}. 
This graph represents the most important result of the present work.

\section{Conclusions}
The recent scientific literature has explored in several ways the relevance of stochastic resonance in the functioning of the auditory system \cite{schilling2021stochastic, Yashima2021, IJASEIT9438, Schilling-preprint}. In this paper we have shown how the mechanism of stochastic resonance coupled with a high-pass filter may hint at a straightforward -- albeit partial -- explanation of the equal loudness curves. As such, the model fulfils the educational goal that we stated in the introduction.

\medskip

\begin{figure}[ht]
  \centering
\includegraphics[width=0.5\textwidth]{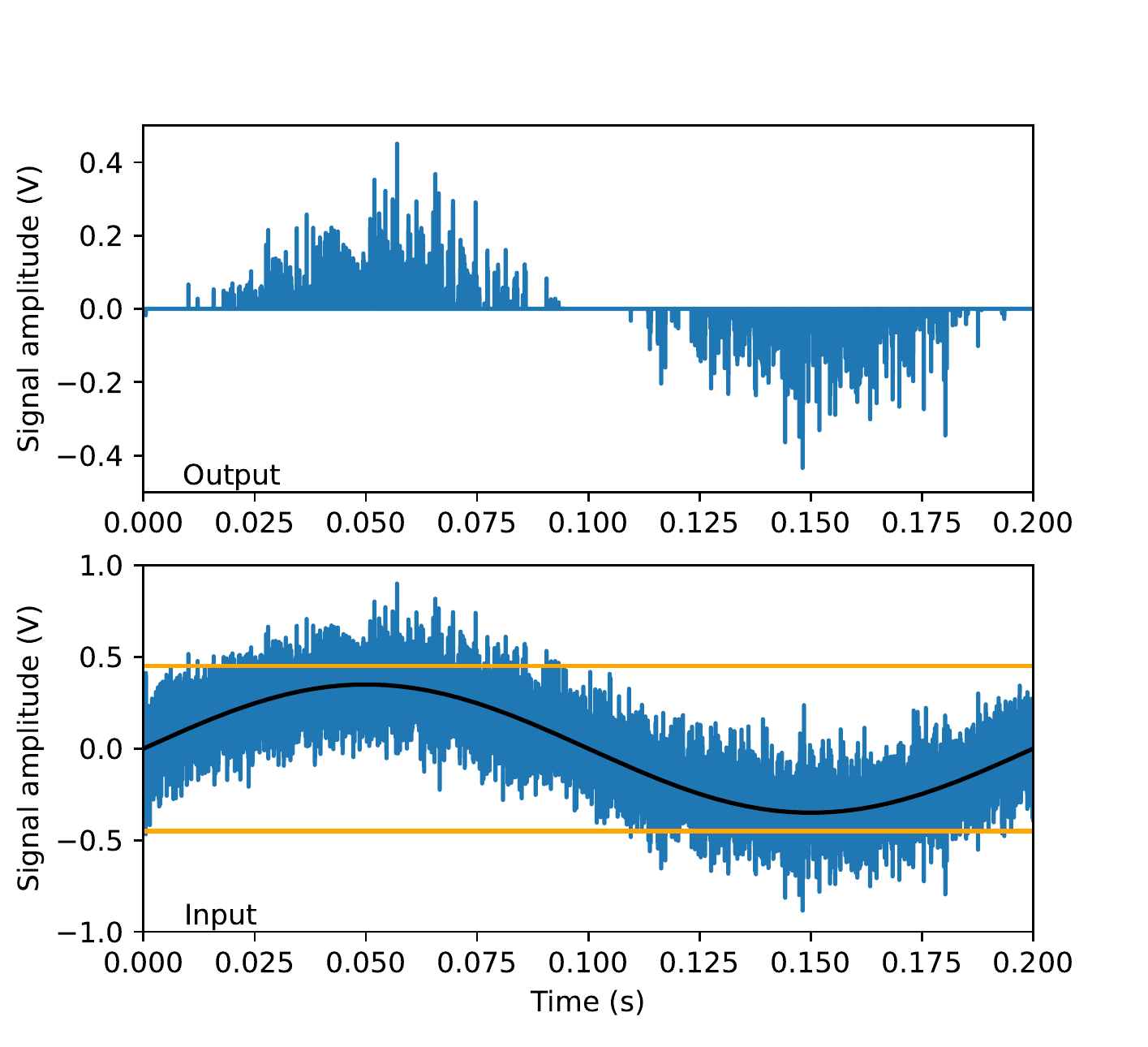}
    \caption{Graphical representation of the LCD system for $f_s=5$~Hz, $\Delta U = 0.45$~V, $D=0.15$~V and $\varepsilon=0.35$~V.
    }\label{fig:LCD-freqCounter3}
\end{figure}

Although the model presented here is highly conjectural, it can be extended in many ways that can potentially be of interest in a more complex model of human hearing. Consider for instance \Fref{fig:SpikeTrain}: in the case of a white background noise, the number of noise spikes that pass the threshold is a Poisson process with a mean that depends on the root-mean-square (RMS) noise amplitude and on the threshold value. By adding a sinusoidal signal like in \Fref{fig:LCD-4000Hz}, we notice that the rate is slightly higher whenever a peak (either positive or negative) occurs. This behavior becomes more prominent for low-frequency deterministic signals, as shown in figure \Fref{fig:LCD-freqCounter3}, where we see that for a given threshold-crossing rate associated with a specific RMS noise amplitude -- threshold value combination we could infer the period of the sine wave by counting the individual positive or negative pulses. Thus, with additional logical circuitry, this simple threshold detector could measure the dominant instantaneous frequency in a signal, greatly extending its reach.

\section*{ORCID iDs}
Francesco Veronesi \texttt{https://orcid.org/0000-0002-6603-2561}\\
Edoardo Milotti \texttt{https://orcid.org/0000-0001-7348-9765}

\section*{References}
\bibliography{references}

\end{document}